\documentclass[sigconf]{acmart}

\usepackage{graphicx}
\usepackage{adjustbox}
\usepackage{tabularx}
\usepackage{subcaption}
\usepackage[figuresright]{rotating}
\usepackage{makecell}
\usepackage{array}
\usepackage{natbib}
\usepackage{caption}
\usepackage{multirow}
\usepackage{url}
\usepackage{float}
\usepackage{hyperref}
\usepackage{adjustbox}
\usepackage{threeparttable}
\usepackage{rotating} 
\usepackage{array}
\usepackage{makecell}
\usepackage{tikz}
\usepackage{afterpage}
\usepackage{enumitem}
\usepackage{color, colortbl}
\usepackage{makecell}
\usepackage{wrapfig}
\usepackage[linesnumbered]{algorithm2e}
\definecolor{Gray}{gray}{0.9}
\definecolor{LightGray}{gray}{0.6}
\usepackage[most]{tcolorbox}
\hypersetup{colorlinks, citecolor={black}, filecolor=black, linkcolor=black, urlcolor=black}

\usepackage{xcolor}
\definecolor{green(munsell)}{rgb}{0.0, 0.66, 0.47}
\definecolor{cadmiumgreen}{rgb}{0.0, 0.42, 0.24}
\definecolor{cobalt}{rgb}{0.0, 0.28, 0.67}
\definecolor{amber(sae/ece)}{rgb}{1.0, 0.49, 0.0}
\newlength\MAX  

\newlength\BARSIZE  

\newcommand*\ChartBarBlue[1]{\textcolor{cobalt}{\rule{\BARSIZE}{2ex}}}
\newcommand*\ChartBarGreen[1]{\textcolor{green(munsell)}{\rule{\BARSIZE}{2ex}}}
\newcommand*\ChartBarOrange[1]{\textcolor{amber(sae/ece)}{\rule{\BARSIZE}{2ex}}}

\AtBeginDocument{%
  \providecommand\BibTeX{{%
    \normalfont B\kern-0.5em{\scshape i\kern-0.25em b}\kern-0.8em\TeX}}}

\copyrightyear{2023}
\acmYear{2023}
\setcopyright{acmlicensed}\acmConference[ICMI '23]{INTERNATIONAL CONFERENCE ON MULTIMODAL INTERACTION}{October 9--13, 2023}{Paris, France}
\acmBooktitle{International Conference on Multimodal Interaction (ICMI '23), October 9--13, 2023, Paris, France}
\acmPrice{15.00}
\acmDOI{10.1145/3577190.3614129}
\acmISBN{979-8-4007-0055-2/23/10}

\begin{document}

\newcommand{\lakmal}[1]{\textcolor{black}{#1}}
\newcommand{\nathan}[1]{\textcolor{orange}{#1}}

\title{Understanding the Social Context of Eating with Multimodal Smartphone Sensing: The Role of Country Diversity}

\author{Nathan Kammoun}
\orcid{0009-0000-4712-7277}
\affiliation{%
  \institution{Idiap Research Institute \& EPFL}
  \country{Switzerland}}

\author{Lakmal Meegahapola}
\orcid{0000-0002-5275-6585}
\affiliation{%
  \institution{Idiap Research Institute \& EPFL}
  \country{Switzerland}
}

\author{Daniel Gatica-Perez}
\orcid{0000-0001-5488-2182}
\affiliation{%
  \institution{Idiap Research Institute \& EPFL}
  \country{Switzerland}
}

\renewcommand{\shortauthors}{Kammoun et al.}

\begin{abstract}
Understanding the social context of eating is crucial for promoting healthy eating behaviors. Multimodal smartphone sensor data could provide valuable insights into eating behavior, particularly in mobile food diaries and mobile health apps. However, research on the social context of eating with smartphone sensor data is limited, despite extensive studies in nutrition and behavioral science. Moreover, the impact of country differences on the social context of eating, as measured by multimodal phone sensor data and self-reports, remains under-explored. To address this research gap, our study focuses on a dataset of approximately 24K self-reports on eating events provided by 678 college students in eight countries to investigate the country diversity that emerges from smartphone sensors during eating events for different social contexts (alone or with others). Our analysis revealed that while some smartphone usage features during eating events were similar across countries, others exhibited unique trends in each country. We further studied how user and country-specific factors impact social context inference by developing machine learning models with population-level (non-personalized) and hybrid (partially personalized) experimental setups. We showed that models based on the hybrid approach achieve AUC scores up to 0.75 with XGBoost models. These findings emphasize the importance of considering country differences in building and deploying machine learning models to minimize biases and improve generalization across different populations.
\end{abstract}

\ccsdesc[500]{Human-centered computing~Empirical studies in ubiquitous and mobile computing}

\keywords{smartphone sensing, passive sensing, multimodal sensor data, eating behavior, social context, mobile food diary, mobile health}

\maketitle

\section{Introduction}
The eating behavior of college students has significant implications for their overall well-being. As a result, behavioral and nutrition scientists have conducted extensive research on the interplay between food consumption and health \cite{schwerin1982food, meegahapola2020protecting}. To this end, prior research has emphasized that eating is a holistic event with interconnected dimensions such as the type of food, context, and time, among other aspects \cite{bisogni_2}. The social context of eating is a factor that shapes people's food consumption, along with other contextual cues such as mood and location \cite{nutrition_science_social_context, social_modeling_of_eating, meegahapola2020alone}. While the social context is multidimensional, prior research has conceptualized eating alone or with others as two basic types of social contexts. Therefore, several studies have investigated the impact of eating alone versus eating with others on food consumption \cite{nutrition_science_alone_or_not_1, nutrition_science_alone_or_not_2}. Moreover, the social context of eating is associated with key aspects of eating episodes, such as mood, physical condition, time, and location.

"Food and Nutrition" is among the most common categories of mobile health applications, but most dietary pattern assessment methodologies rely on Ecological Momentary Assessments (EMA) and survey questionnaires \cite{bruening2016mobile}. However, measuring eating behavior with such techniques is hardly ever done consistently, as self-reports can be burdensome for users \cite{jung2020foundations}. Moreover, typical dietary assessment surveys do not always capture the social context and other contextual cues associated with an individual's food consumption \cite{dietary_assessment, dietary_assessment_2}. As an alternative, longitudinal behavior modeling with multimodal sensing can address the challenges of dietary assessment methods by capturing the interplay between food consumption and contextual characteristics while replacing burdensome questionnaires \cite{one_more_bite, bits_and_bites}. Inference models relying on features derived using phone sensors (activity types, step count, location, app usage, typing and touch events, screen on and off episodes, etc.) can be used to understand key contextual aspects affecting food consumption \cite{food_mood_context, alcool_alone_with_others, meegahapola2020alone}. However, the social context of eating, especially in the case of multiple countries, has not systematically been studied with mobile technologies.

Diversifying the data for training models by considering more countries is fundamental to improving the performance of the models and their ability to generalize to a broader range of populations. Generalizing research results by building diversity-aware models can help overcome issues related to biases and reproducibility and aim toward better generalizability of mobile sensing research across different contexts \cite{bangamuarachchi2023inferring, generalization_across_countries, nanchen2023keep, schelenz2021theory}. However, even though computer vision and natural language processing research have studied cross-country or cross-dataset generalization of models, mobile sensing studies focused on longitudinal behavior modeling have started doing so only recently \cite{Xu2023Globem, Meegahapola2023Generalization, generalization_across_countries}. Considering these aspects, this paper aims to answer the following research questions:

\noindent \textbf{RQ1:} What behavioral and contextual characteristics can be extracted from analyzing the social context surrounding eating events among college students in eight countries, using multimodal smartphone sensing data and self-reports?

\noindent \textbf{RQ2:} Can multimodal phone sensor data be utilized to infer the social context of eating events, and if so, how does the geographical diversity affect the inference performance?

In answering these research questions, we provide the following contributions. 

\noindent \textbf{Contribution 1:} We conducted a study using a large dataset (24K eating event self-reports and corresponding multimodal sensor data from ten modalities) collected from college students in eight countries (UK, Italy, Denmark, Paraguay, Mongolia, Mexico, India, and China) with the goal of identifying behavioral differences among countries. We first conducted a descriptive analysis that demonstrated the natural time-dependency of eating patterns and then examined the hourly distribution of eating-alone versus eating-with-others events across the eight countries to provide initial evidence of cross-country diversity. We found that the hourly pattern of the eating social context varies across countries as both the time of eating and the social context are tied to social practices. Furthermore, we statistically analyzed the interplay between smartphone features and the social context during eating events to demonstrate possible differences across countries. We found that while some features are used consistently across all settings, others are only relevant in some countries.

\noindent \textbf{Contribution 2:} Using smartphone features and binary target data on the social context of eating (alone or with others), we developed inference models with various methods. We trained a number of model configurations: \textit{i)} separate models for each country; \textit{ii)} aggregate models based on geographical proximity, such as Asia, Latin America, and Europe, and \textit{iii)} a model that pools data from all countries. For each case, we evaluated population-level (non-personalized) and hybrid (partially personalized) models using AUC as the performance metric. The experimental results showed that the AUC of population-level models was 0.58, while the hybrid models achieved AUC scores of up to 0.75. We also analyzed the feature importance extracted from the hybrid models to obtain further insight into the cross-country differences of inference models. By displaying all feature importances based on their common use across all countries, we show that the same subset of features is used consistently across all countries with varying degrees of relevance. These findings provide a comprehensive understanding of country diversity by revealing more complex relationships in the data that could not be identified through statistical analysis alone.

\section{Background and Related Work}
{\textit{2.1\hspace{0.2 cm}Eating as a Holistic Event.}} Studies in nutrition and behavioral sciences have shown that eating behavior is influenced by various factors, including situational and behavioral ones. Bisogni et al. \cite{bisogni_2} proposed a contextual framework for eating and drinking events, which includes eight interconnected dimensions such as ``food and drink, time, location, activities, social context, mental processes, physical condition, and recurrence''. Following this framework, eating is considered a complex event that is influenced by a range of internal and external factors, including the social context. Researchers have highlighted the social context of meals as a fundamental aspect of food consumption \cite{social_modeling_of_eating, nutrition_science_social_context}. The social context itself comprises several dimensions, such as the type of relationships or the number of participants. Specifically, whether an individual is alone or with others has often been used in behavioral studies as two key categories of social context \cite{ICMI_commensality_mfd, nutrition_science_alone_or_not_2}. Prior research in nutrition and behavioral sciences has demonstrated that eating in highly social contexts is likely to influence the amount of food consumed. The presence of one or more people during meals can also lead to impression management or social facilitation, resulting in under-eating and overeating, respectively \cite{nutrition_alone_overeating, social_modeling_of_eating}. Therefore, the longitudinal understanding of social context during eating events could enable timely feedback and interventions for individuals who use mobile health applications.

\noindent {\textit{2.2 \hspace{0.2 cm}Multimodal Mobile Sensing for Eating Behavior Modeling.}} The advancement of mobile sensing technologies has allowed researchers to monitor eating behaviors in real-time using wearable sensors to detect eating events or characterize behavioral and contextual attributes around eating episodes. Such developments in ubiquitous computing have provided researchers with additional insight into eating behaviors \cite{ICMI_eating_analysis, Bin_Morshed_2020, EarBit, NeckSense, bangamuarachchi2022sensing, meegahapola2020smartphone}. Prior research has emphasized the importance of studying the context surrounding eating events to better understand an individual's eating social context \cite{food_mood_context, ICMI_social_dimension_eating}. However, most of these studies have primarily used self-reports instead of sensor data. Moreover, the datasets used in these studies were not collected from multiple countries, limiting the ability to investigate country differences, as seen through both self-reports and sensor data. In studies that use sensing for studying eating behavior, Biel et al. \cite{bits_and_bites} demonstrated that sensor data from phones could be used to distinguish between meal and snack events. Meegahapola et al. \cite{one_more_bite} showed that overeating events could be inferred with multimodal sensor data with personalized accuracy of up to 87\%. Similarly, another study \cite{alcool_alone_with_others} used sensor data to infer the social context during drinking events with accuracies in the range of 0.75-0.86. However, despite being briefly discussed in one previous study \cite{meegahapola2020alone}, no other study has extensively examined the social context of eating with sensor data collected from multiple countries using the same study protocol, thus allowing for intuitive comparisons.

\noindent {\textit{2.3 \hspace{0.2 cm} Diversity Awareness.}} Previous studies in the field of machine learning have highlighted the need to ensure model accuracy and fairness across diverse populations \cite{diversity_framework}. The concept of data diversity has been applied to various domains, including computer vision and natural language processing \cite{shankar2017classification}. However, applying data diversification to mobile sensing is challenging due to the lack of large-scale multimodal datasets collected using a consistent protocol across multiple countries. Therefore, it is crucial to develop diversity-aware approaches to machine learning-based modeling of sensor data to examine model generalization across countries. Mobile sensing data can exhibit diversity due to various factors such as user behavior, differences in devices, and mobile networks in the country of interest. Smartphone usage can also vary significantly between countries, even within a given population, such as students. While being used consistently in mobile sensing studies without considering generalization to other countries, passive sensing features from WiFi or location need to be considered with care in deployment settings \cite{generalization_across_countries}. Recently, several research groups have explored this direction of research by collecting multimodal sensing datasets for multiple countries for longitudinal behavior modeling \cite{Meegahapola2023Generalization, Khwaja_2019}. These studies have investigated the effect of distributional shifts across countries. However, studies focusing on social context inference during specific events such as eating and drinking have only used data from one or two countries \cite{alcool_alone_with_others, labhart2021ten}. Therefore, no comparable multimodal smartphone sensing datasets exist for similar analysis, as previous work relied on data collected with contrasting protocols and features, hence restricting comparisons across countries \cite{meegahapola2020alone}. Therefore, the impact of country diversity on social context inference during eating events, or eating events in general, has not been thoroughly discussed and requires further investigation.

\section{Data, Features and Target Classes}
\subsection{Dataset Information} To examine the research questions, we rely on a dataset used in our previous work  \cite{Meegahapola2023Generalization, giunchiglia2022aworldwide}. This dataset was collected during an in-the-wild study that was conducted over a four-week period in November 2020, aimed to explore the everyday behavior of college students through multimodal mobile phone sensors. Deployed in eight countries across various regions of the world with the same protocol, the study aimed to investigate the effect of country diversity on key aspects of mobile sensing. 

\begin{table}[]
\caption{Mobile Sensing Data Collection Summary.}
\label{participants}
\resizebox{\columnwidth}{!}{%
\begin{tabular}{lrrrrr}
\rowcolor[HTML]{EFEFEF} 
\textbf{Country (N)} & \textbf{$\mu$ Age ($\sigma$)} & \textbf{\% Women} & \textbf{\# Self Reports} & \textbf{\# Eating Reports} & \textbf{\% Eating Alone} \\ 
China (41)  & 26.2 (4.2) & 51 & 30,406  & 1548   & 45 \\
Denmark (24)  & 30.2 (6.3) & 58 & 12,354  & 613    & 41 \\
India (39)  & 23.7 (3.2) & 53 & 4478    & 340    & 42 \\
Italy (240) & 24.1 (3.3) & 58 & 176,135 & 12,697 & 28 \\
Mexico (20)  & 24.1 (5.3) & 55 & 11,908  & 773    & 40 \\
Mongolia (214) & 22.0 (3.1) & 65 & 121,809 & 5,674  & 15 \\
Paraguay (28)  & 25.3 (5.1) & 60 & 11,790  & 729    & 29 \\
UK (72)  & 26.6 (5.0) & 66 & 31,989  & 1,599  & 51 \\ 
\rowcolor[HTML]{EFEFEF} 
\textbf{Total} (678) & 24.2 (4.2) & 58 & 400,829 & 23,973 & 36  
\end{tabular}%
}

\end{table}

The first data collection phase gathered demographic general data about each participant. The second phase involved collecting data through a smartphone application. Participants were asked to fill out time diaries throughout the day, indicating what they were doing (from 34 activities), location (from 26 semantic locations), social context (from 8 configurations), and mood (valence with a five-point scale). 

In this study, we focus on eating events from the larger dataset, which contains reports for a wide range of activities. Therefore, we only considered the samples for which the reported activity is eating, which drastically reduced the number of data points, with eating events representing 5\% to 7\% of the data, depending on the country. Table~\ref{participants} presents statistics about the participants, including the total number of reports and the number of eating reports, the sample sizes are fairly uneven between countries ranging from 4478 to 176,135 reports. The participants were asked to choose from a variety of categories when filling out self-reports; possible answers for social context include: alone, with relatives, with friends, with classmates, and so on. These answers were translated into a binary target Alone or With Others in line with previous studies {\cite{nutrition_science_alone_or_not_1, nutrition_science_alone_or_not_2, alcool_alone_with_others}}. Table~\ref{participants} also provides the class distributions for each country.

\noindent \subsection{Smartphone Sensing Features} The app continuously collected data from more than thirty smartphone sensors, divided into two categories: continuous sensing modalities (activity type, step count, location, phone signal, WiFi, Bluetooth, battery, and proximity) and interaction sensing modalities (notifications clicking, application usage, screen episodes, and user presence). A time-window-based approach was used to process the feature data. All sensor measurements were aggregated with self-reports to create features characterizing the time window during which the report occurred. The length of the window was a ten-minute window around eating event self-report, during which statistics about each smartphone modality were computed \cite{Meegahapola2023Generalization}, obtaining over 100 features (more details in the appendix). Two additional features were computed from timestamps: a binary feature stating whether the current day is a weekday or weekend and the hour of report collection. Further, in the context of smartphone sensing, missing data can occur for multiple reasons: device in low-consumption mode, sensor failure, user privacy settings, airplane mode, or simply the type of phone used, not including necessary hardware. To deal with this issue, the feature modalities for which more than 70\% of the data was missing were dropped, namely: Bluetooth low energy, Bluetooth normal, Cellular GSM, and Cellular WCDMA. While most of the tools used in this study were robust against missing values, an imputed version of the dataset was still needed in some cases. A k-Nearest Neighbor (k-NN) imputation {\cite{knn-imputation}} was used to impute missing values from the remaining features in training sets, whenever needed.

\begin{figure}[tb]
  \centering
  \includegraphics[width=0.9\linewidth]{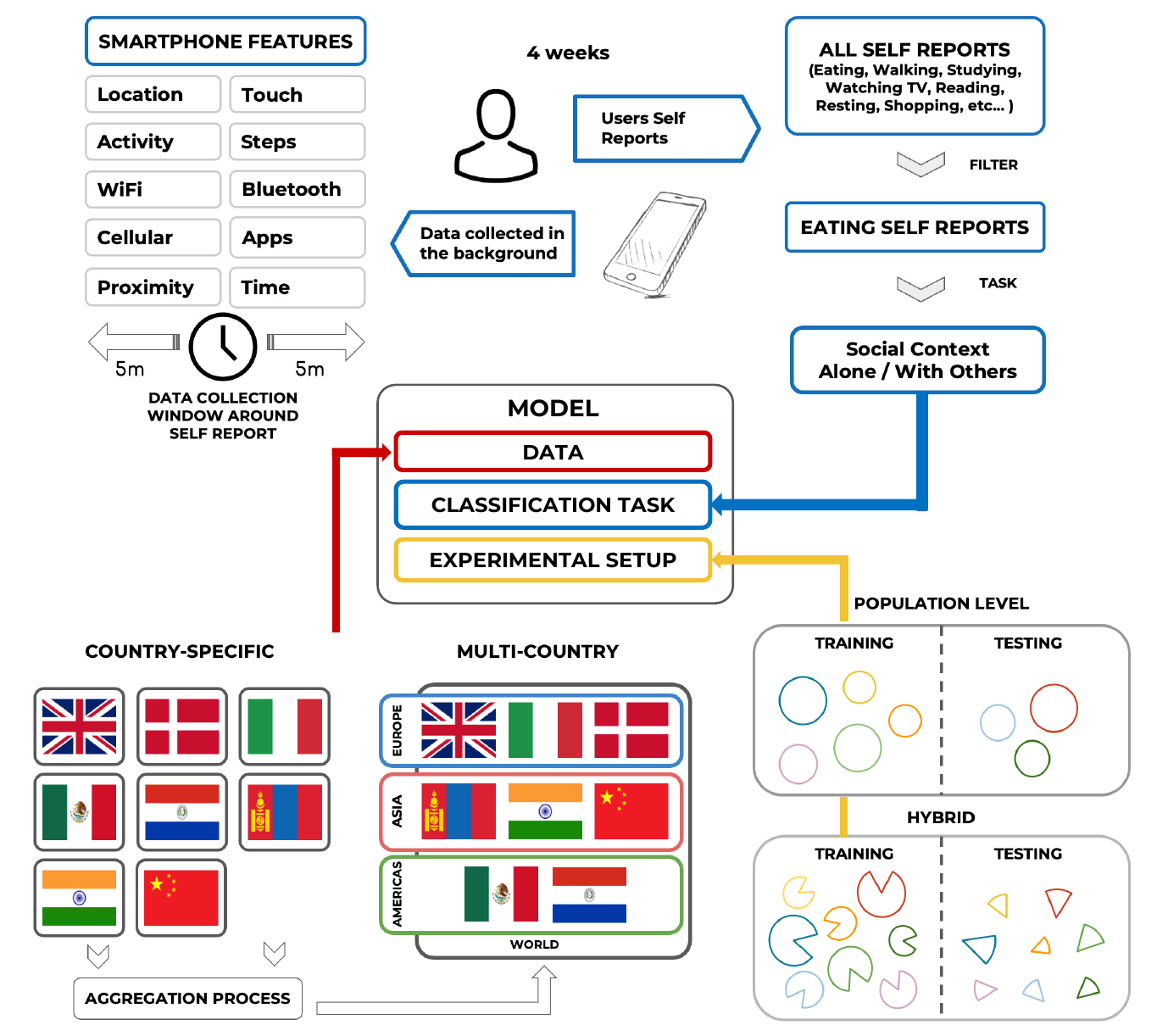}
  \caption{High-Level Overview of the Study}
  \label{diagram}
\end{figure}

\section{Behavior \& Context Around Eating Social Context (RQ1)}
\subsection{Methodology}

To answer the first research question, a descriptive and statistical analysis was conducted in three separate components. First, we analyzed the hourly distribution of eating events with respect to the social context using self-reports. Next, we performed a statistical analysis to explore the relationship between smartphone sensor features and the two social contexts of eating. Lastly, we created a two-dimensional embedding for the extracted sensor features to visually showcase the cross-country differences to better understand the relationship between sensor data and eating social contexts.

\subsubsection{Descriptive Analysis}

As eating events are inherently time-dependent, we analyzed the hourly distribution of eating reports for all countries. We provided two separate density plots to investigate how eating events were distributed during the day for different social contexts (alone vs. with others) in Figure~\ref{fig:hourly_distribution}. We then examined the hourly distribution of eating events to identify the relationship between time and social context across countries.

\subsubsection{Statistical Analysis} 
To understand the cross-country differences, the class distributions of each eating context outcome were statistically analyzed to uncover features that define eating social context in each country. Hence, t-tests were performed to understand class distributions with mobile sensing features. The top five features with the highest t-statistic {\cite{student1908probable}} and p-values lesser than 0.05 were reported in Table~\ref{tab:table-stats} alongside the effect size (Cohen's-d) {\cite{cohen_1988}} \footnote{As a rule of thumb, a value below 0.2: small effect size, a value of 0.5: medium effect size, and a value of 0.8 or higher: large effect size.}. In order to gain insight into feature distributions across countries, in Figure~\ref{fig:stats_context}, we plotted the effect sizes with features grouped by sensor types such as location, WiFi, cellular data, notifications, proximity, applications, activity, steps, and touch. The purpose of this plot is to display which feature groups are the most important and to provide a means of effectively analyzing differences between countries.

\subsubsection{Visualizing Diversity} 

We also present a method for visualizing individual self-reports in the high-dimensional feature space. The t-Stochastic Neighbor Embedding (t-SNE) {\cite{JMLR:v9:vandermaaten08a}} algorithm was used for this purpose. t-SNE is a dimensionality reduction technique that maps high-dimensional features to a two-dimensional space for visualization purposes. First, t-SNE constructs a probability distribution of high-dimensional features by assigning higher probabilities to similar data points. Next, it defines a similar probability distribution over points in the low-dimensional space and optimizes the embedding by minimizing the Kullback-Leibler (KL) divergence between both distributions. In summary, t-SNE maps each eating event in the high-dimensional feature space to a two-dimensional point such that similar events are mapped to nearby points, and dissimilar events are mapped to distant points with high probability. This method allows us to gain insight into individual self-reports and their relationships in the feature space. In order to treat all dimensions with equal importance when performing the dimensionality reduction, we re-scaled features with the Z-score.

\noindent \textit{{Country Specific Embedding.}} A separate t-SNE analysis was conducted on each country's dataset to obtain a country-specific feature space mapping. This analysis aimed to show how user-specific eating events are sensed by plotting each user's eating events in a different color. The first step was to standardize the data, as t-SNE relies on a measure of similarity between data points. The embedding was then fitted using the hyper-parameters: $perplexity=30.0$, $early\_exageration=12.0$, and $learning\_rate=200.0$. However, for most countries, the number of users was too high to plot all the sensed eating events for each user, which would have resulted in a more difficult reading. So, for each country, only twenty users were randomly selected for plotting after fitting the embedding on the entire country's dataset. Results are presented in Figure~\ref{fig:t-SNE_users}.

\noindent \textit{{Multi-Country Embedding.}} A single t-SNE was also applied on the whole dataset, comprising an aggregation of all countries, to investigate country-specific behaviors by assigning a color to each country and plotting the reports for each country in separate plots for clarity. To avoid allowing the country with more data to imbalance the stochastic embedding, a balanced aggregation of countries was created by randomly selecting a subset of reports in each country. Then, data were standardized, and the embedding was fitted using the same parameters as before. While this approach employs individual points for each report to reveal clusters, overlapping points can make it challenging to observe how eating events are distributed in the 2D plots. To compensate for this, an additional density plot was presented where the 2D space was partitioned based on polar coordinates, and the density of events in each partition was calculated, as given in Figure~\ref{fig:t-SNE_countries} and Figure~\ref{fig:t-SNE_darts}.

\subsection{Results}

\subsubsection{Descriptive Analysis}

\begin{figure}
     \centering
     \begin{subfigure}[b]{0.235\textwidth}
         \centering
         \includegraphics[width=\textwidth]{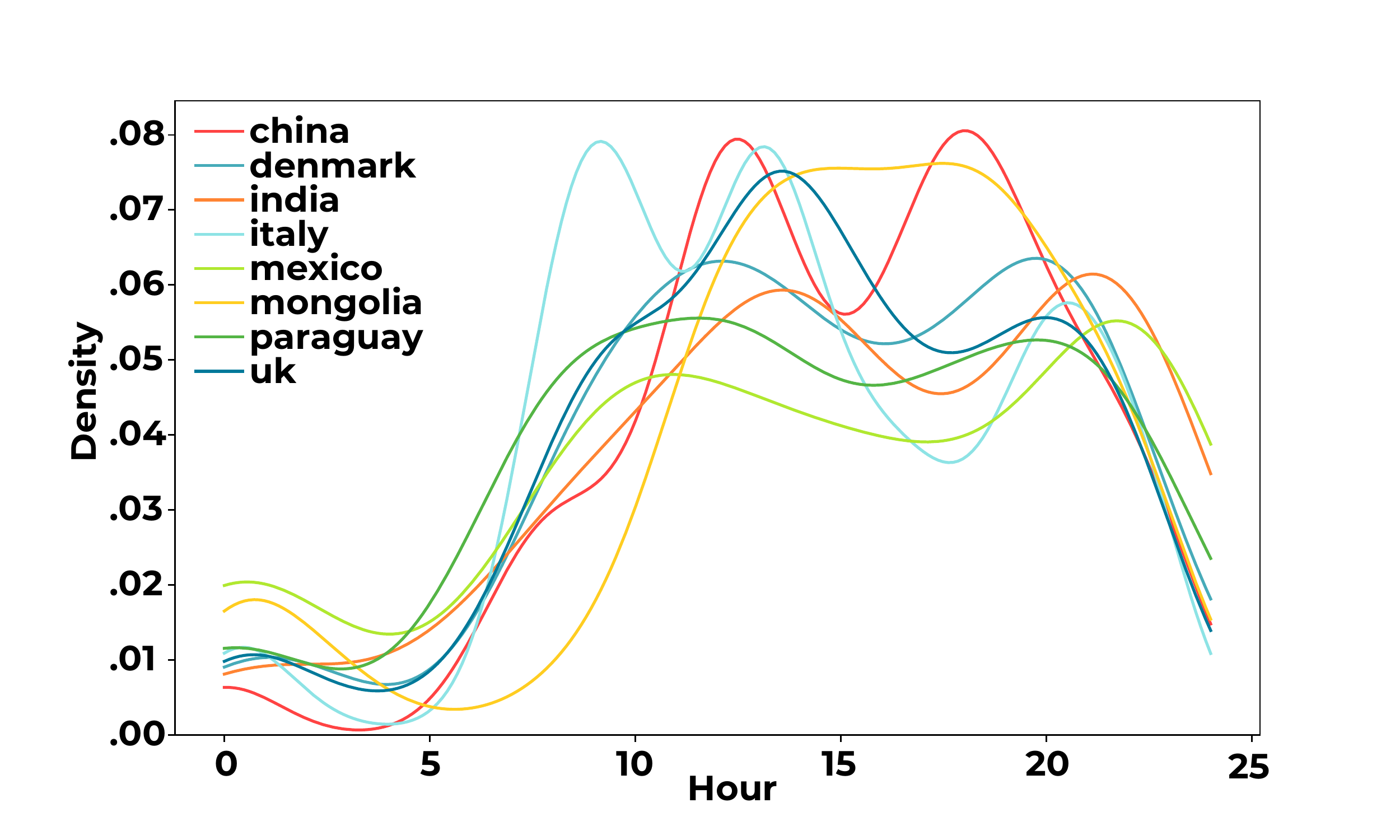}
         \caption{Alone}
         \label{eating_alone}
     \end{subfigure}
     \hfill
     \begin{subfigure}[b]{0.235\textwidth}
         \centering
         \includegraphics[width=\textwidth]{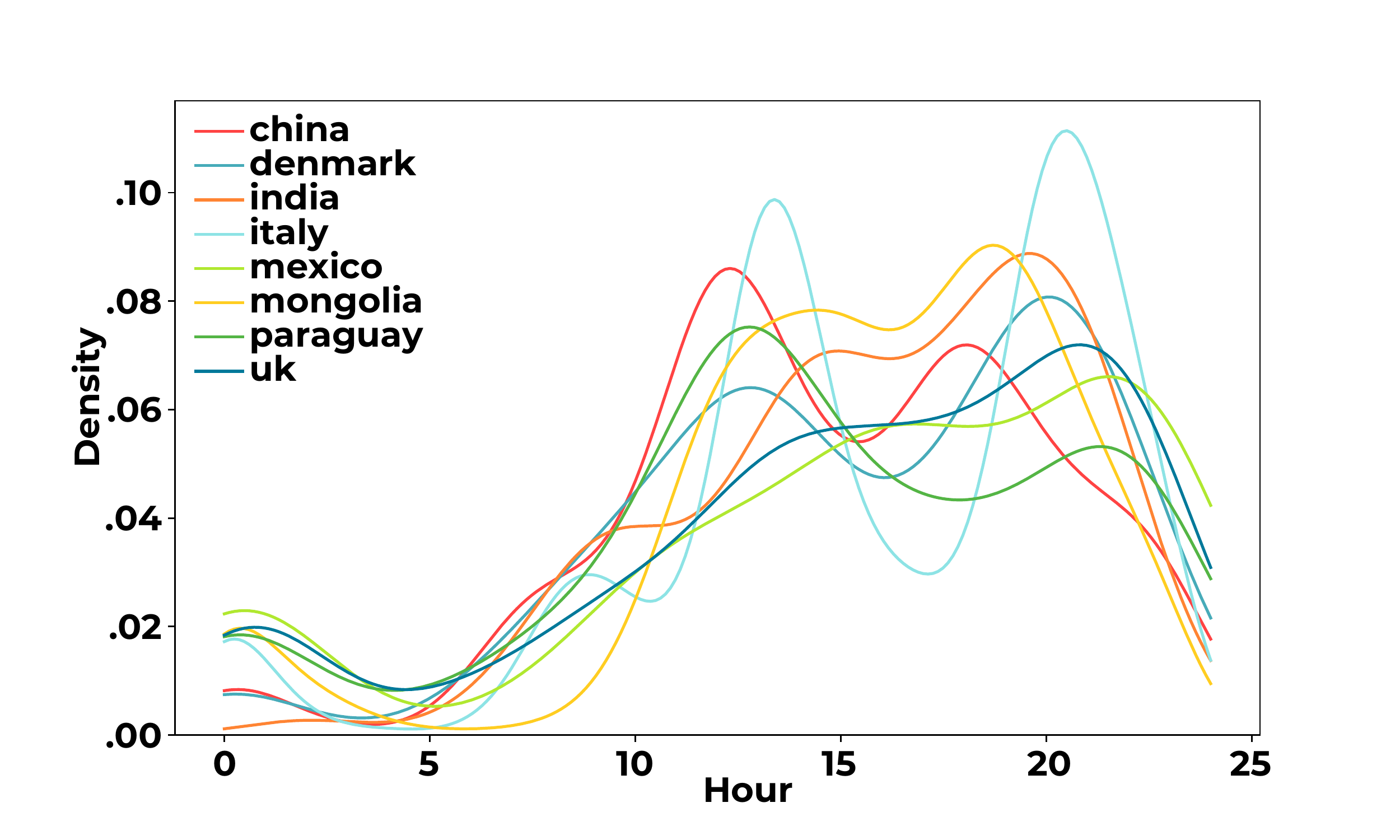}
         \caption{With Others}
         \label{eating_w_others}
     \end{subfigure}
     \caption{Hourly Distribution of Eating Events}
     \label{fig:hourly_distribution}
\end{figure}

Figure~\ref{eating_alone} and Figure~\ref{eating_w_others} display the hourly distributions of eating alone and with others events. As these outcomes are measured during eating events, they naturally follow the trend of the general eating schedule. It is worth noting that as the classes are imbalanced, the dominant class pattern is always close to the general eating schedule. Therefore, it is more interesting to explore the hourly distribution of the minority class. These plots enable us to better understand the eating context's time dependency across countries. In Figure~\ref{eating_alone}, we can observe that all countries show a higher distribution of `alone' reports during the morning in comparison with the `with others' distribution, suggesting that breakfast might have been taken alone. This finding is consistent with previous research in nutrition sciences, showing that eating alone most commonly takes place in the morning and midday \cite{nutrition_science_alone_or_not_1}. Please note that the data was collected during the covid19 pandemic year of 2020, so this influences typical socialization patterns. Moreover, the peak of `alone' reports during the evening for Italy significantly decreases, indicating that dinner is more likely to happen with other people. So, the eating context is time-dependent, and while this holds for all countries with varying degrees, the patterns are not the same, which provides additional evidence of the cross-country diversity regarding eating behaviors.

\subsubsection{Statistical Analysis with Hypothesis Testing.}

\begin{table}[]
\caption{t-statistic (TS) and Cohen’s-d (CD). TS with p-values above 0.05 and CD with a 95\% confidence interval overlapping with 0 are marked with a star (*). p-values are reported after the Bonferroni correction.}
\label{tab:table-stats}
\resizebox{\columnwidth}{!}{%
\begin{tabular}{llll llll}
\rowcolor[HTML]{EDEDED} 
 \textbf{Country} & \textbf{Feature} & \textbf{TS} & \textbf{CD} & \textbf{Country} & \textbf{Feature} & \textbf{TS} & \textbf{CD} \\ 
\multicolumn{1}{c}{} & proximity mean & 3.15 & 0.19 &  & app tools & 3.09 & 0.23 \\
\multicolumn{1}{c}{} & app tools & 2.48 & 0.13 &  & location speed mean & 2.60 & 0.17 \\
\multicolumn{1}{c}{} & app health \& fitness & 1.97 & 0.10* &  & location radius of gyration & 2.58 & 0.17 \\
\multicolumn{1}{c}{} & location\_speed\_max & 1.80* & 0.09* &  & weekday & 2.56 & 0.18 \\
\multicolumn{1}{c}{\multirow{-5}{*}{China}} & app\_not\_found & 1.68* & 0.13* & \multirow{-5}{*}{Mexico} & hour & 2.42 & 0.18 \\ \arrayrulecolor{Gray}
    \midrule
 & app tools & 5.20 & 0.46 &  & screen time total & 6.25 & 0.24 \\
 & weekday & 4.77 & 0.38 &  & screen max episodes & 5.77 & 0.22 \\
 & cellular lte std & 3.52 & 0.35 &  & screen \# episodes & 5.69 & 0.20 \\
 & app communication & 3.16 & 0.28 &  & touch \# events & 5.15 & 0.17 \\
\multirow{-5}{*}{Denmark} & activity tilting & 3.14 & 0.34 & \multirow{-5}{*}{Mongolia} & screen time / episode & 4.83 & 0.19 \\ \arrayrulecolor{Gray}
    \midrule
 & wifi max rssi & 9.01 & 1.36 &  & weekday & 3.63 & 0.28 \\
 & wifi mean rssi & 8.56 & 1.29 &  & app tools & 2.64 & 0.22 \\
 & wifi std rssi & 7.27 & 1.13 &  & location\_altitude\_min & 2.32 & 0.21 \\
 & location altitude min & 7.24 & 1.01 &  & app board & 2.15 & 0.13* \\
\multirow{-5}{*}{India} & proximity std & 3.24 & 0.5 & \multirow{-5}{*}{Paraguay} & app strategy & 1.90* & 0.12* \\ \hline
 & hour & 19.02 & 0.37 &  & app tools & 6.06 & 0.32 \\
 & weekday & 9.26 & 0.18 &  & hour & 3.40 & 0.17 \\
 & app tools & 8.54 & 0.17 &  & activity invehicle & 3.39 & 0.22 \\
 & wifi std rssi & 5.94 & 0.14 &  & activity onbicycle & 3.36 & 0.22 \\
\multirow{-5}{*}{Italy} & wifi max rssi & 5.77 & 0.14 & \multirow{-5}{*}{UK} & activity running & 3.33 & 0.21 \\ \bottomrule
\end{tabular}%
}
\end{table}

The results of the hypothesis testing are presented in Table \ref{tab:table-stats}. While some features are consistently used across countries, the vast majority of them differ, highlighting the diversity in smartphone usage. Time-related features rank among the most relevant for all countries except for China, Mexico, and Paraguay, confirming the hourly distributions' findings. Notably, the weekday feature is visible among features for Denmark, Italy, Mexico, and Paraguay, indicating a different social context depending on whether it is a weekend or a weekday. WiFi features play a significant role in India, which could be because, in the university where the study was conducted, students tend to gather in places with WiFi network-dense areas to eat. The most relevant features seem to point towards a measure of phone usage. For example, in Mongolia, screen features indicate an increase in smartphone usage, which could mean that the user is alone when eating. The presence of proximity features in China is also noteworthy. The proximity sensor is a good indicator of whether a phone is in the user’s pocket/bag or hand, which could have a relation with being with people. It is important to note that the results are not all easily interpretable, and the table only shows a selection of features. However, the variety of features across countries provides a good display of cross-country diversity in the sensor data, for the two eating social contexts.

\begin{figure}[tb]
  \centering
  \includegraphics[width=\linewidth]{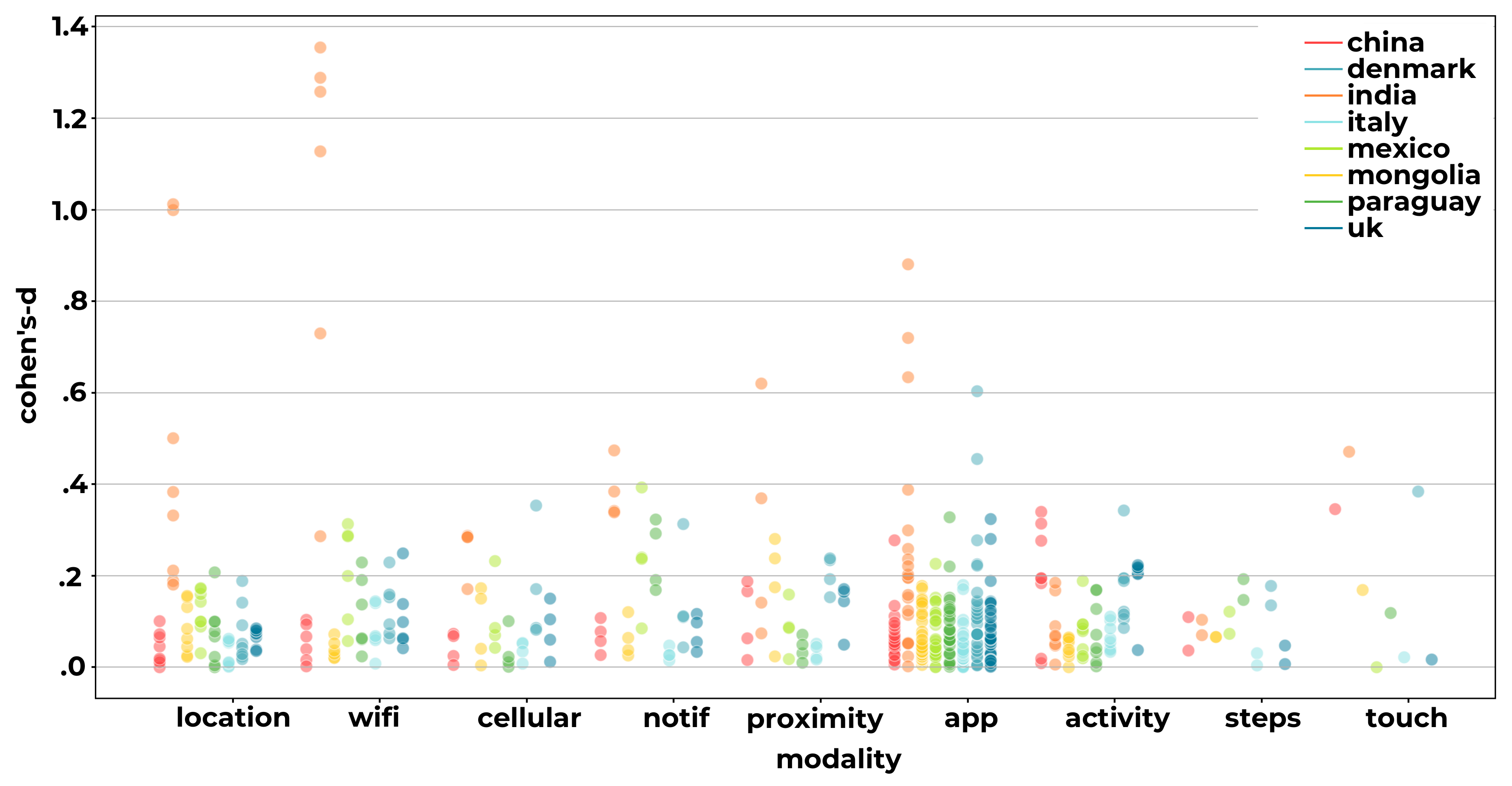}
  \caption{Effect Sizes for Social Context of Eating}
  \label{fig:stats_context}
\end{figure}

\subsubsection{Statistical Analysis on Effect Sizes and Feature Groups.} Figure~\ref{fig:stats_context} presents the plot for effect sizes. It is observed that most features have an effect size below 0.25, but some features have a large effect size, and their Cohen's-d 95\% interval does not overlap with zero, indicating higher reliability in results. The results show that India has above-average effect sizes on most features, except for activity-type features. Larger effect sizes were found for activity features in China and UK, while Paraguay has low Cohen's-d for cellular and proximity features. Mongolia also had smaller effect sizes for WiFi and notifications compared to other countries. These differences may be attributed to cultural aspects or mobile networks and hardware. In conclusion, analyzing effect sizes among feature groups can help identify differences between countries and discover which features could be useful for inference.

\subsubsection{Visualizing Diversity with Country Specific Embeddings.}

\begin{figure}[tb]
  \centering
  \includegraphics[width=\linewidth]{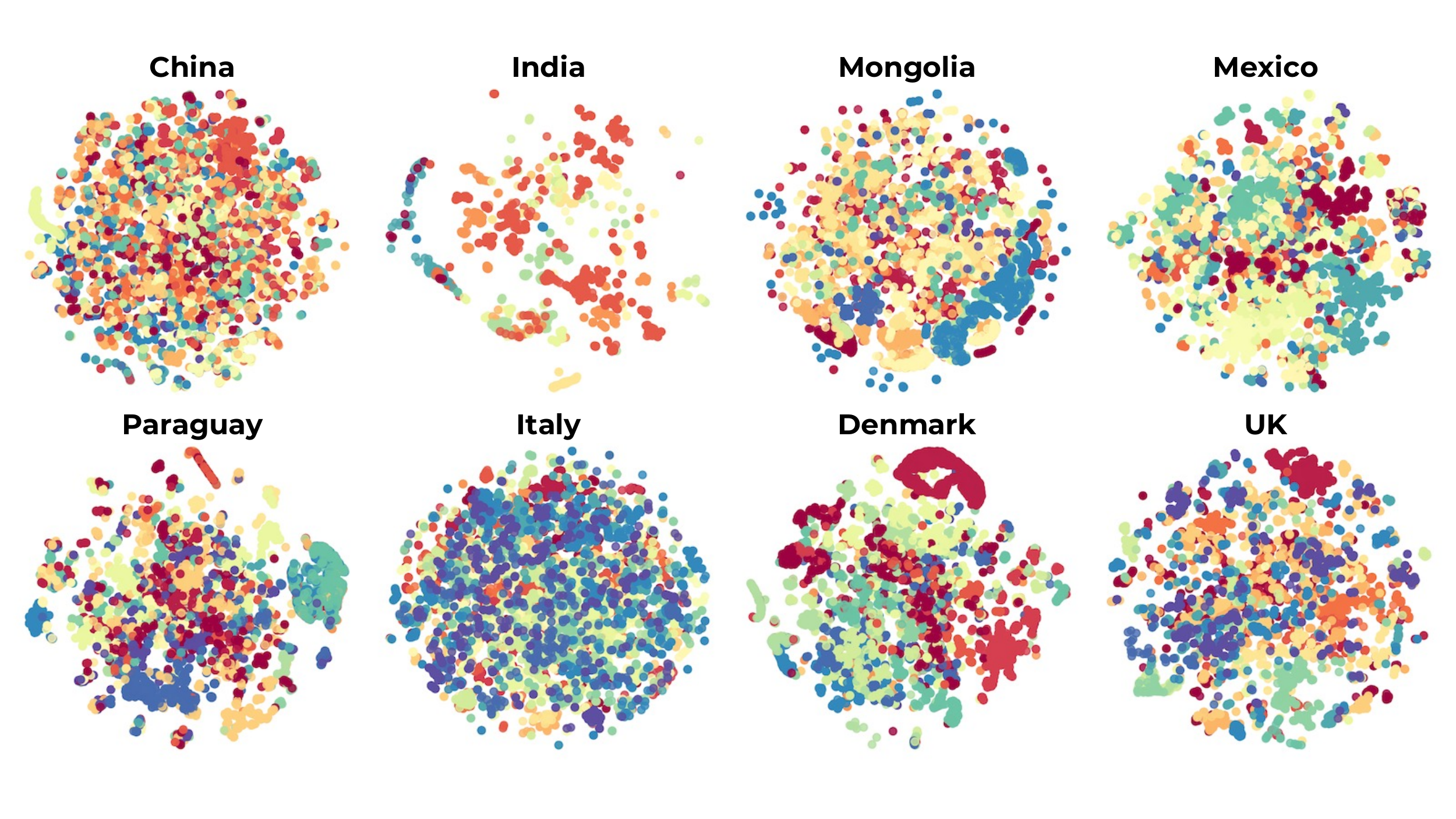}
  \caption{Country-Specific t-SNE: each color stands for a user.}
  \label{fig:t-SNE_users}
\end{figure}

Figure~\ref{fig:t-SNE_users} shows the outcomes of the mappings, indicating that sensed eating events are highly personalized for each user. The presence of clusters of points of the same color across all countries indicates that user behavior is distinct and that phone features could be highly associated with the user's habits. However, some countries, such as China, exhibit less distinctive clusters, suggesting that sensor features during eating events are more similar across users in that country. The presence of clusters of the same color at different locations on the plot could indicate that the same user's behavior varies across different contexts. Despite the loss of information resulting from mapping a feature space from 96 dimensions to 2 dimensions, these plots demonstrate the variability across users in the same country and how that variability applies in different parts of the world.

\subsubsection{Visualizing Diversity with Multi-Country Embedding.} 
\begin{figure}[tb]
  \centering
  \includegraphics[width=\linewidth]{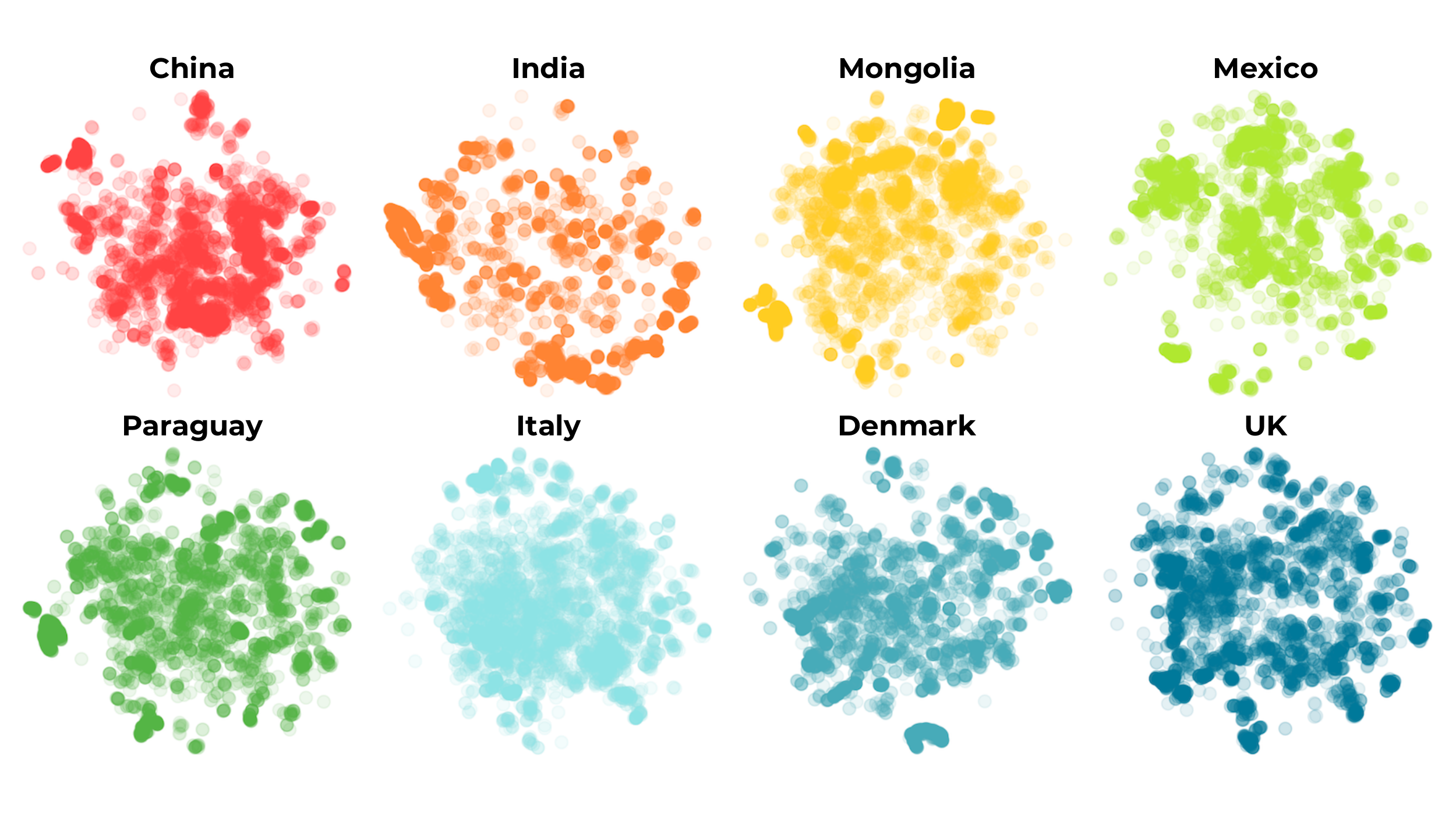}
  \caption{t-SNE fitted on a worldwide aggregation of countries, each country's eating events are plotted separately}
  \label{fig:t-SNE_countries}
\end{figure}

\begin{figure}[tb]
  \centering
  \includegraphics[width=\linewidth]{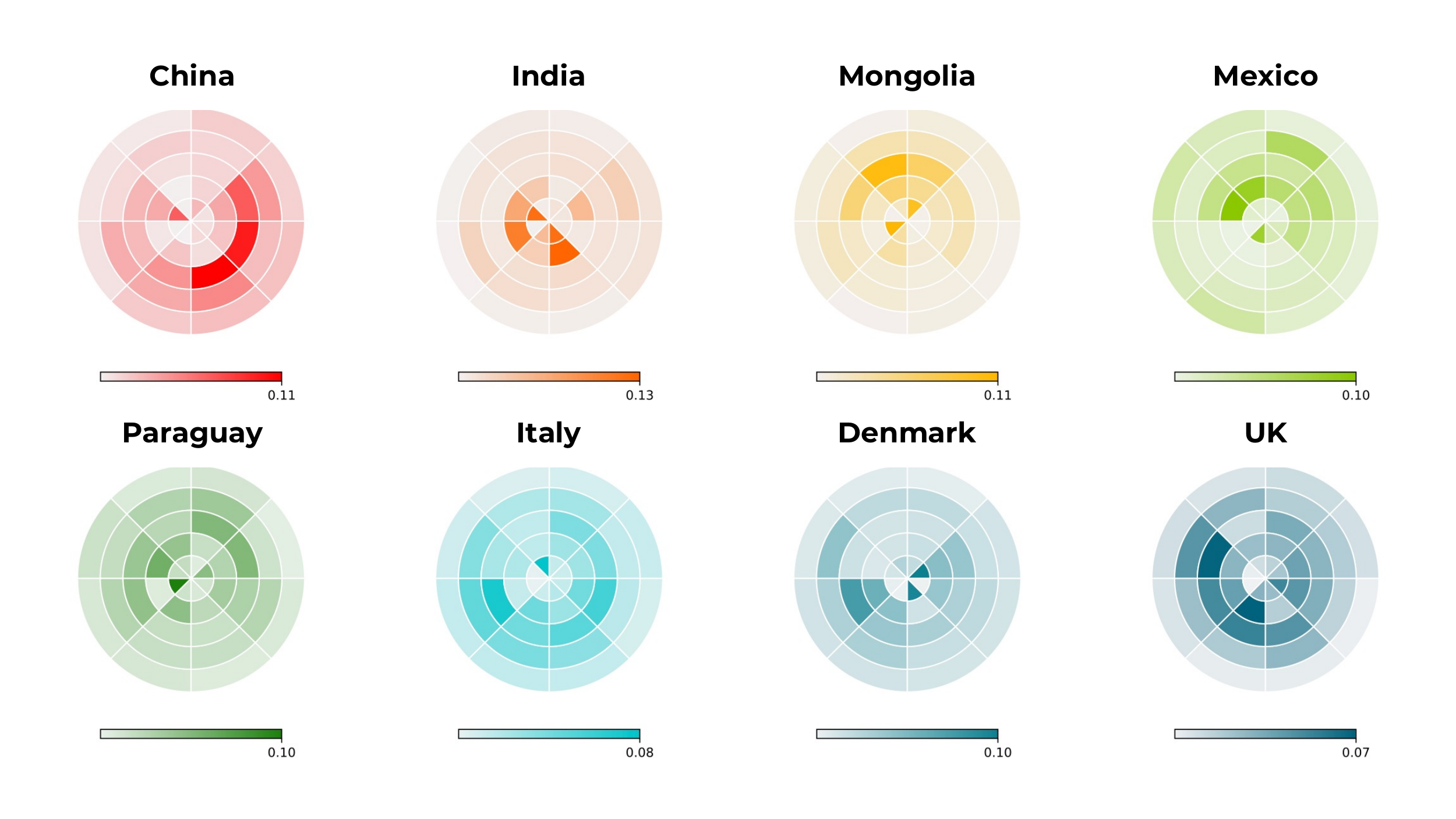}
  \caption{t-SNE fitted on the aggregation of all countries, the country-specific density of eating-reports in the 2D space}
  \label{fig:t-SNE_darts}
\end{figure}

Figure~\ref{fig:t-SNE_countries} demonstrates the clustering of sensed eating events from each country in different areas of the plot, indicating the high degree of country-specificity. This highlights the necessity of developing diversity-aware models that take into account the cross-cultural variability in smartphone usage and eating behavior. The proximity of the points in the low-dimensional space represents the similarity between the sensed eating events. Although comparing distributions in two dimensions has its limitations, some general observations can still be made. For instance, in Figure~\ref{fig:t-SNE_darts}, the clustering of points for Denmark, the UK, and Italy suggests that these European countries have similar phone usage patterns during eating events. In contrast, the sensed eating events from China, India, and Mongolia are located in opposite areas of the low-dimensional space, indicating differences in smartphone use and cultural practices.

\section{Inferring the Social Context of Eating (RQ2)}
\subsection{Methodology}

The objective of this section of our study is to assess the capability of eating social context inference models to generalize across users and countries. We used smartphone sensor features collected within a time interval around eating events as the input space and the social context of eating as the target variable. We analyzed country-specific models as well as multi-country models. A series of experiments were carried out to compare the use of different inference methods such as multi-layer perceptron neural networks (MLPs), Random Forests, and Support Vector Classifiers. eXtreme Gradient Boosting (XGboost), a regularizing gradient boosting framework, {\cite{xgboost}} was chosen based on better performance and ease of adaptability and tuning across experimental setups.

Despite having observed a significant time-dependency in the eating behaviors, it was decided not to train separate models, e.g., weekend/weekdays, due to lack of data in weekends. However, models trained on time-related features only were used as a baseline to assess the added benefits of using smartphone sensing in comparison with a naive prediction solely based on time features. Finally, with a sample size ranging from 340 (India) to 12,667 (Italy) eating reports, and given the class imbalance, special care was given to addressing these issues when creating aggregations, performing data splits, and evaluating the model performance.

\textit{Population-Level models} are based on the leave-k-participants-out cross-validation {\cite{crossval}}, which involves splitting data into training and testing sets based on individual users, creating disjoint sets between the two. The splitting is stratified to ensure that the percentage of Alone and With Others samples is maintained in both sets. This allows for testing the model's robustness against user-specific behaviors, simulating a scenario where the model is trained on a population, and new users join the system.

\textit{Hybrid models} are based on K-fold cross-validation {\cite{crossval}}, and this approach involves randomly splitting the data into training and testing sets without considering individual users. The testing set contains a fraction of the eating social context reports from each user, and the splitting is done in a stratified manner to preserve the percentage of Alone and With Others samples in each split. This approach simulates a scenario where the model is trained on a population, and the same participants provide additional reports to the system, resulting in partially personalized models.

In all inference tasks, an interesting property of XGBoost was utilized, which is the sparsity-aware split finding technique \cite{Chen:2016:XST:2939672.2939785}. This technique allows training on sparse data by letting tree branches handle missing values. Initially, models were trained with both imputed and sparse data. However, it was observed that the results were slightly better when using the sparse method. Therefore, the sparse method was chosen for all further training and testing.

The hyper-parameters of XGBoost have a large impact on its performance and generalizability. To optimize these parameters, nested cross-validation was used. The hyper-parameters considered were the step size shrinkage (\textit{eta}), the minimum loss reduction required to make a split (\textit{gamma}), the maximum tree depth (\textit{max tree depth}), and the minimum child weight (\textit{min child weight}). Grid search was used to find the best combination of hyper-parameters. Five train/test folds were used to evaluate the models' performance. Each train set was further divided into five nested cross-validation splits to determine the best parameters. The final hyper-parameters were selected by choosing the most frequently picked parameters among the five folds. To validate the performance of the models, a random binary vector was generated and used as a baseline to compare against the model's prediction. In addition to the time-based baseline, this baseline was used to verify that the prediction outperformed a random guess.

The countries were aggregated based on their continental region, clearly being aware of this oversimplifying assumption of similarity. Thus, Europe was formed by combining Italy, the UK, and Denmark; Latin America by combining Paraguay and Mexico; and Asia by combining Mongolia, India, and China. All countries were then aggregated to form a multi-country world dataset. An equal number of users were selected from each country to ensure that the number of eating reports in each country equaled the number of reports in the country with the least reported eating events. Stratified group K fold splits were performed in each separate country to perform the cross-validation splits for population-level models before aggregating them into one training and one testing set. The same procedure was applied to hybrid models to maintain an equal amount of data from each country in train and test sets. To ensure the generalizability of the results on country aggregations, selected users for the majority of countries were resampled at each repetition.

The chosen metric for evaluating the results is the Area Under the Receiver Operating Characteristic (AUC). Due to the limited amount of data available, it was not feasible to perform more than five folds on the countries with less data. However, evaluating models on just five folds can lead to uneven performances between each run. Therefore, repeated stratified group K folds and repeated stratified K folds were conducted for the population level and hybrid experimental setups, respectively. The number of repetitions was set to 5, and the folds were randomized each time to obtain different data splits. To summarize, the models were evaluated on five repetitions of 5-fold validation, and the model performances were averaged over the resulting 25 folds. One of the objectives of building inference models is to leverage the feature importance as additional information about cross-country diversity. The feature importance coefficients were also averaged over the 25 folds.

\subsection{Results}

\begin{table}[]
\caption{Mean ($\overline{A}$) and Standard Deviation ($A_{\sigma}$) of inference AUC for population-level and hybrid models, calculated with XGBoost and Baseline (trained on time based features)}
\label{tab:table-merged}
\resizebox{0.9\columnwidth}{!}{%
\begin{tabular}{llll lll}
 &
  \multicolumn{3}{c}{\cellcolor[HTML]{9FBAC1}\textbf{Population Level}} &
  \multicolumn{3}{c}{\cellcolor[HTML]{B1CBD2}\textbf{Hybrid}} \\
 &
  Model &
  Baseline &
  Random &
  Model &
  Baseline &
  Random \\ \arrayrulecolor{Gray}
    \midrule
China    & .48 (.05) & .47 (.03) & 0.5 & .68 (.02) & .50 (.02) & 0.5 \\
Denmark  & .52 (.08) & .50 (.06) & 0.5 & .74 (.04) & .57 (.03) & 0.5 \\
India    & .50 (.04) & .49 (.03) & 0.5 & .75 (.06) & .58 (.05) & 0.5 \\
Italy    & .58 (.01) & .57 (.02) & 0.5 & .64 (.01) & .57 (.01) & 0.5 \\
Mexico   & .53 (.05) & .51 (.06) & 0.5 & .65 (.03) & .57 (.01) & 0.5 \\
Mongolia & .49 (.01) & .50 (.01) & 0.5 & .57 (.01) & .50 (.01) & 0.5 \\
Paraguay & .49 (.06) & .55 (.05) & 0.5 & .61 (.04) & .55 (.03) & 0.5 \\
UK       & .55 (.04) & .53 (.03) & 0.5 & .75 (.04) & .56 (.02) & 0.5 \\ \hline
Latin America & .53 (.02) & .51 (.04) & 0.5 & .63 (.03) & .54 (.02) & 0.5 \\
Asia     & .51 (.06) & .52 (.04) & 0.5 & .72 (.03) & .55 (.04) & 0.5 \\
Europe   & .56 (.05) & .54 (.03) & 0.5 & .72 (.04) & .56 (.02) & 0.5 \\
World    & .53 (.03) & .53 (.02) & 0.5 & .69 (.02) & .53 (.02) & 0.5 \\ \hline 
\end{tabular}%
}
\end{table}%

\subsubsection{{Population Level Results.}} In general, the country-specific models trained with the population-level approach were unsatisfactory, as shown in Table~\ref{tab:table-merged}. The models exhibited uneven performance across countries, where social context seems more easily inferred in European countries than in Asia and Latin America. For instance, the models achieved AUC scores of 0.52 (0.08) for Denmark, 0.55 (0.04) for the UK, 0.58 (0.01) for Italy, and 0.56 (0.05) for Europe aggregate. The lower model variance observed in Italy could be explained by the high number of users, which provided the model with a more accurate picture of the population, leading to a comparatively robust model. Conversely, Denmark exhibited high model variance (0.08), indicating that the model either performed well or poorly, depending on the user split. This situation arises when the social context is reflected differently across users, causing the model to learn a trend that does not always apply to the remaining users in the test set, depending on the split. This finding is consistent with the tight clusters observed for Denmark in Figure \ref{fig:t-SNE_users}, pointing towards a variety of phone sensor features. Interestingly, the world model provided good results comparatively, whereas Asia and Latin America consistently outperformed the performance of their respective countries. This could be attributed to the increase in training samples being creating robust models to counter the diversity in data. However, this conclusion should not hold in the case where countries have sufficient data individually.

\subsubsection{{Hybrid Results.}} The inclusion of a fraction of each user's data in the models resulted in improvements in the results. For instance, the models for China, Mongolia, and Paraguay improved from showing no predictive power to showing reasonable results, with AUCs of 0.68 (0.02), 0.57 (0.01), and 0.61 (0.04), respectively. Generally, the country-specific models showed reasonable performance with an AUC of 0.75 (0.06) for India, 0.75 (0.04) for the UK, and 0.74 (0.04) for Denmark. The aggregations also provided good results with AUC of 0.72  (0.03), 0.72 (0.04), and 0.69 (0.02) for Asia, Europe, and World, respectively. These findings support the conclusions made for population-level setups. However, Latin America suffered from the country's aggregation. In summary, the hybrid models performed well and consistently outperformed the baseline based on time features. The improvements observed with hybrid model results are in line with the diversity of user practices observed in Figure \ref{fig:t-SNE_users}.

\begin{figure}[tb]
  \centering
  \includegraphics[width=\linewidth]{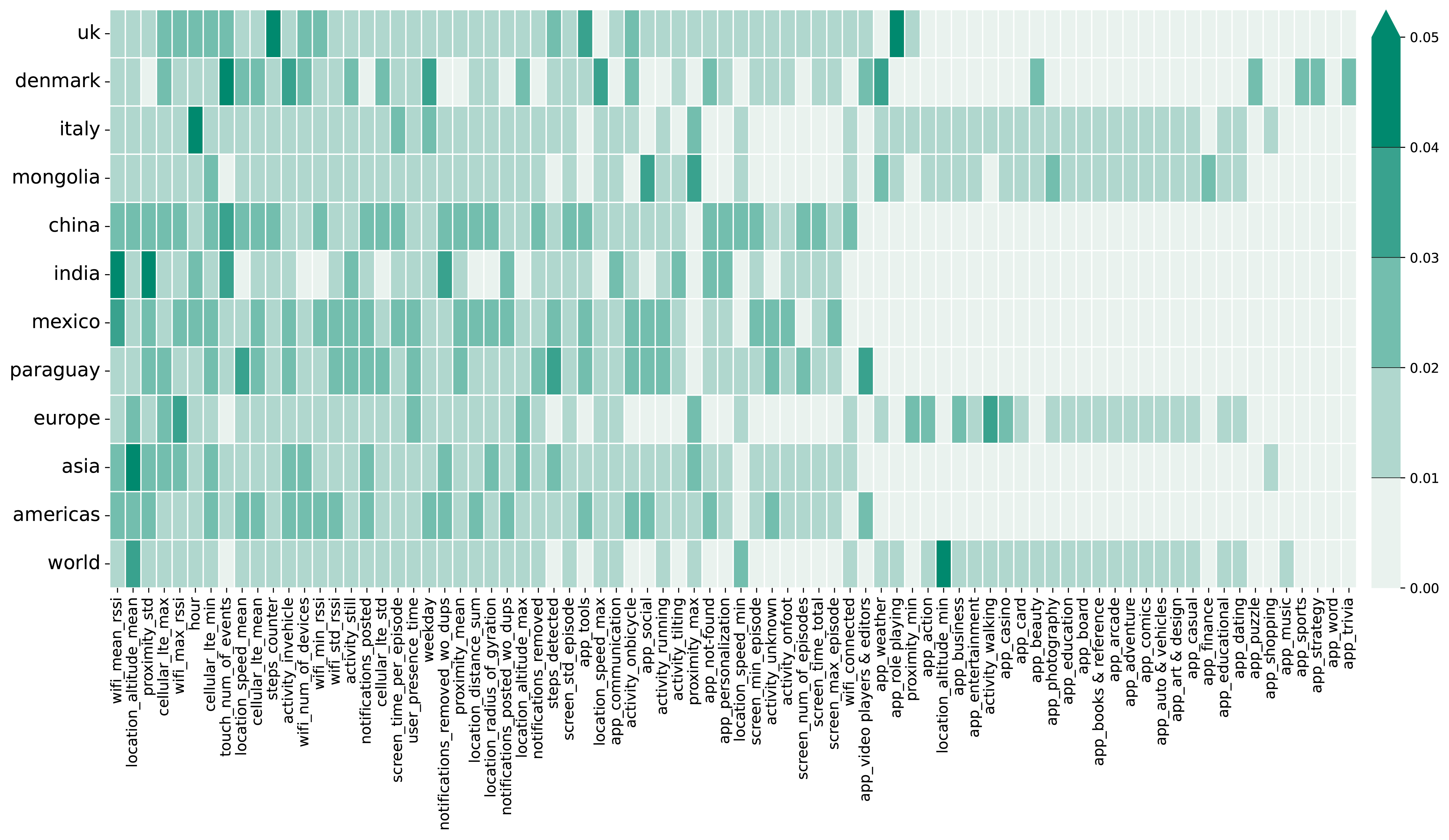}
  \caption{XGBoost Feature Importance Values}
  \label{gini_context}
  \vspace{-0.2 in}
\end{figure}

\noindent  \subsubsection{{Feature Importance.}} The use of XGBoost results in a loss of the natural explanatory power of models like Random Forests, but it is still possible to extract feature importances. XGBoost outputs a weight measure that reflects the number of times a feature is used to split the data across all trees. After extracting the feature importances at each evaluation split and averaging over the 25 folds, the values are displayed in Figure \ref{gini_context} for hybrid models. The features are ranked left to right, based on the mean across all configurations, to visualize better which features are shared across countries and continents. Higher feature importances in countries aggregations are not necessarily ones used in any of the separate countries, which leads us to think that feature importance cannot be approached as the sum of individual ones in a cross-country context. India and Asia had very strong feature importance on their most important features, which imbalanced the color scale and reduced the contrast in the rest of the plot. Therefore, the color scale was limited to a certain threshold visible at the top of the color bar, which means all values beyond the threshold take a single color. It is worth noting that the countries with the most data, Italy and Mongolia, have a more spread-out use of features. The overall fading of the plots towards less important features shows that the different models tend to be using the same features, and the clear and dark tiles that stand out highlight features that are specific to one country. In summary, the feature importances resonate with the effect sizes listed in Table~\ref{tab:table-stats} with the difference that the model has the ability to leverage complex relations in the data that may not be visible otherwise.

\section{Discussion}
\subsection{Implications}

This study provides insights into the contextual characteristics associated with eating events in different countries, which can aid in understanding the differences between eating events from a mobile sensing perspective. Our findings highlight the need to account for country biases when building machine learning models based on mobile sensing data, as these biases can arise from differences in demographic attributes, lifestyle, culture, and other factors. Further, the study shows how features associate with the social context of eating events in different countries and shows the importance of country diversity in how these events are sensed. The identification of different user practices associated with eating events in different countries provides useful insights for future data collection and highlights the impact of country diversity on the sensing of eating events. Therefore, we recommend that researchers consider relevant contextual factors and country differences during the planning phase of their studies for training models with multimodal data that captures a range of situational and behavioral contexts.

Our study also reveals that some countries exhibit a degree of similarity in terms of sensed eating events, but geographical proximity cannot be used as a proxy for similar user behaviors. While clear country-wise patterns and clusters arise when analyzing the countries, the difference in scores for population-level versus hybrid models, as well as the clusters observed in user behavior, highlight the fact that social context differs greatly across users, even in the same country. Moreover, the amount of user diversity among countries is subject to variation, with user practices being more or less spread out, and researchers should pay specific attention to countries showing a higher level of diversity among users when gathering data or designing studies. Finally, our study demonstrates that feature importance can be used to extract insights into user practices associated with eating events, highlighting the importance of accounting for the complex relations in the data that might not be visible otherwise.

\subsection{Limitations and Future Work}

This study focuses on the impact of geographical diversity on the eating social context, but it acknowledges that diversity is a multifaceted concept that encompasses more than just geographical shifts. The results suggest that assuming similar cultural and social behavior from people living in the same country only holds to a certain extent and that considering more granular aspects of diversity, such as socioeconomic background or cultural norms, could improve inference models' accuracy. Future work can focus on these aspects. Further, the use of additional APIs providing information about location, weather, and behavior could be considered in future studies for additional input of contextual information.

The study uses data that were collected at a single university in each country, which has implications on the representativeness of the samples of students that are considered. In addition, the number of participants is uneven between universities leading to a fairly high data imbalance between countries.

In addition, the data were collected during a period coinciding with the surge of the Covid-19 pandemic in Europe, Paraguay, and Mongolia, during which social distancing measures were imposed or recommended. While the pandemic likely influenced and altered participants' behavior and social practices, the study's findings are expected to hold in the future as remote work/study settings become more prevalent.

The analysis is limited to the two-class social context inference task of determining if an individual is alone or with others, while the dataset includes eight social contexts. While the variable offers valuable insight for mobile food diaries, a more granular approach to social context inference could help gain a better understanding of eating events. The decision to use the two-class inference was motivated by class imbalance, and the fact that the variable is well populated across countries, which is not the case for other class social contexts, making a similar analysis across countries more challenging. Further, the mobile sensing data used in this study can reveal sensitive information about the social context of people, which has ethical implications. Therefore, researchers must carefully consider and address ethical issues to maintain user confidentiality and safeguard against any potential ethical issues.

\section{Conclusion}
This study used a mobile sensing dataset and over 24K self-reports from 678 participants in eight countries, collected over a period of four weeks, to investigate how geographic diversity affects the eating social context inference. We performed an analysis and extracted key aspects of the behavioral and contextual differences that emerge from sensing eating events in different countries. We also evaluated country-specific and multi-country approaches trained on multimodal mobile sensing data to infer eating alone vs. with others, with population-level (non-personalized) and hybrid (partially personalized) models. In addition, the study of feature importance across different models provided additional insight into geographical diversity.  Overall, we highlight the potential for mobile sensing-based machine learning models to generalize across geographically diverse settings, for inferring the social context of eating.

\begin{acks}
This work was funded by the European Union’s Horizon 2020 WeNet project, under grant agreement 823783. We thank all the WeNet partner research teams and volunteer participants, who collectively produced the datasets we used.
\end{acks}

\bibliographystyle{ACM-Reference-Format}
\bibliography{8_Citations}

\end{document}